\documentclass[11pt]{article}
\usepackage{moriond,epsfig}
\usepackage{graphicx}

\def\be{\begin{equation}}
\def\ee{\end{equation}}
\def\bea{\begin{eqnarray}}
\def\eea{\end{eqnarray}}

\begin{document}
\title{Prompt photon production in nuclear collisions}

\author{Fran\c{c}ois Arleo}

\address{~\\ Universit\'e de Li\`ege, Institut de Physique B5 \\
Sart Tilman, B-4000 Li\`ege 1, Belgium}

\maketitle\abstracts{
We discuss the energy loss effects on single prompt photon production in heavy-ion collisions at RHIC and LHC energy. The production of correlated photon pairs at the LHC is also considered.}
\section{Introduction}

The production of ``hard'' probes proves extremely successful to characterize the dense QCD medium --~mostly its gluon density~-- formed in high energy heavy-ion collisions. By hard, we mean those processes with associated scales $Q$ much larger than the typical scales of the medium (let it be the temperature of the saturation scale, $Q \gg T, Q_s$). Provided these scales are well separated, it is sensible to study the effects of the latter on the former. Another feature which makes such processes attractive is the validity of perturbative techniques as $Q \gg \Lambda_{{\rm QCD}}$.

The production of jets, or say large $p_{_T}$ hadrons, appears to be a promising tool to study the parton energy loss due to the multiple gluon radiation induced by the dense medium~\cite{wlossreview}. The spectacular attenuation of large $p_{_T}$ $\pi^0$'s observed at RHIC~\cite{Adcox:2001jp} is a hint that such a mechanism may be at work in central Au-Au collisions at $\sqrt{s} = 200$~GeV. On the other hand, we may want to study the Drell-Yan process --~whose lepton pair does not interact with the surrounding medium~-- which could reflect leading-twist shadowing (or possible gluon saturation) in the~nuclei~wavefunctions. 

\begin{center}
\begin{figure}[htbp]\label{fig:components}
\centerline{\psfig{figure=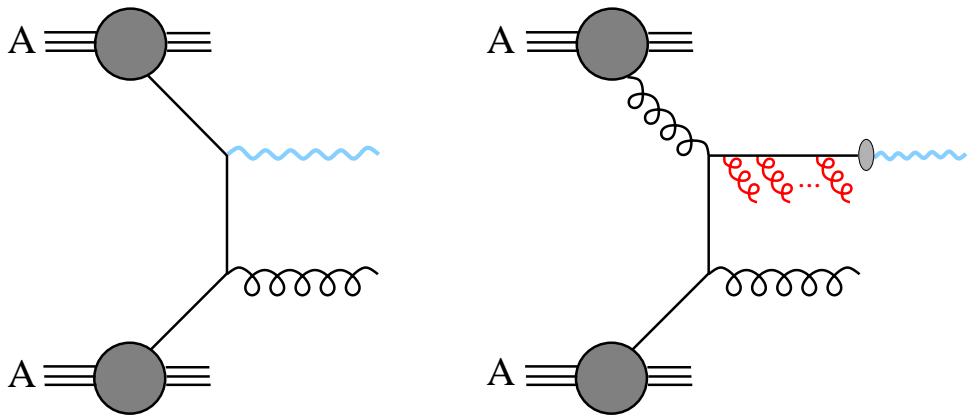,width=12.8cm}}
\caption{Two contributions to prompt photon production at leading-order: direct ({\it left}) and fragmentation ({\it right}) component. The latter may be modified by the soft gluon emission induced by the medium.}
\end{figure}
\end{center}

In this talk, we shall mostly be concerned by prompt photon production~\footnote{We ignore thermal and decay photon production, and drop the word ``prompt'' in the following.} which lies ``in between'' the two above-mentioned processes~\cite{photonreview}. As a matter of fact, two channels may contribute to photon production at leading-order in $\alpha_s$: either a photon takes part directly to the hard subprocess (Figure~\ref{fig:components}, {\it left}), either it comes from the collinear fragmentation of the hard quark or gluon (Figure~\ref{fig:components}, {\it right}). While the former will not be affected by the medium (``Drell-Yan like''), the latter will certainly do (``jet like''). Here, we will first pay attention to the energy loss effects on single photon and pion production. The production of correlated photon pairs at the LHC is then addressed.
\vspace{-0.4cm}
\section{Single photon and pion $p_{_T}$ spectra}

We compute in this section the single photon and neutral pion $p_{_T}$ spectra in $p$-$p$ and A-A collisions to leading-order in $\alpha_s$. Energy loss effects are included in our calculations for both channels by means of medium-modified fragmentation functions~\cite{Wang:1996yh}
\begin{equation}
\label{eq:modelFF}
z\,D_{\gamma,\pi /i}^{med}(z, M, k_{_{T}}) = \int_0^{(1 - z) k_{_{T}}} \, d\epsilon
\,\,{\cal P}(\epsilon, k_{_{T}})\,\,\, z^*\,D_{\gamma,\pi /i}(z^*, M).
\end{equation}
where the scaling variable is shifted from $z$ to $z*$  
\begin{equation}\label{eq:shift}
z = \frac{p_{_{T}}}{k_{_{T}}} \qquad \to \qquad z^* = \frac{p_{_{T}}}{k_{_{T}} - \epsilon} = \frac{z}{1 - \epsilon/k_{_{T}}},
\end{equation}
to account for the energy shift of the leading parton from $k_{_T}$ to $k_{_T} - \epsilon$ with a probability ${\cal P}(\epsilon, k_{_{T}})$. This probability distribution has been computed in Ref.~\cite{Arleo:2002kh} and depends on one typical scale $\omega_c$ related to the gluon content and the spatial extent of the produced medium. For the illustration to come, we shall take $\omega_c = 20$ and 50~GeV for RHIC (Au-Au, $\sqrt{s} = 200$~GeV) and LHC (Pb-Pb, $\sqrt{s} = 5500$~GeV) energy respectively. Although the small $x$ parton densities in nuclei differ somehow to those in a proton, we do not take such shadowing corrections into account in the present qualitative discussion. A more detailed discussion will appear elsewhere~\cite{Arleo:2004xx}.

\begin{center}
\begin{figure}[htbp]\label{fig:rhiclhc}
\centerline{\psfig{figure=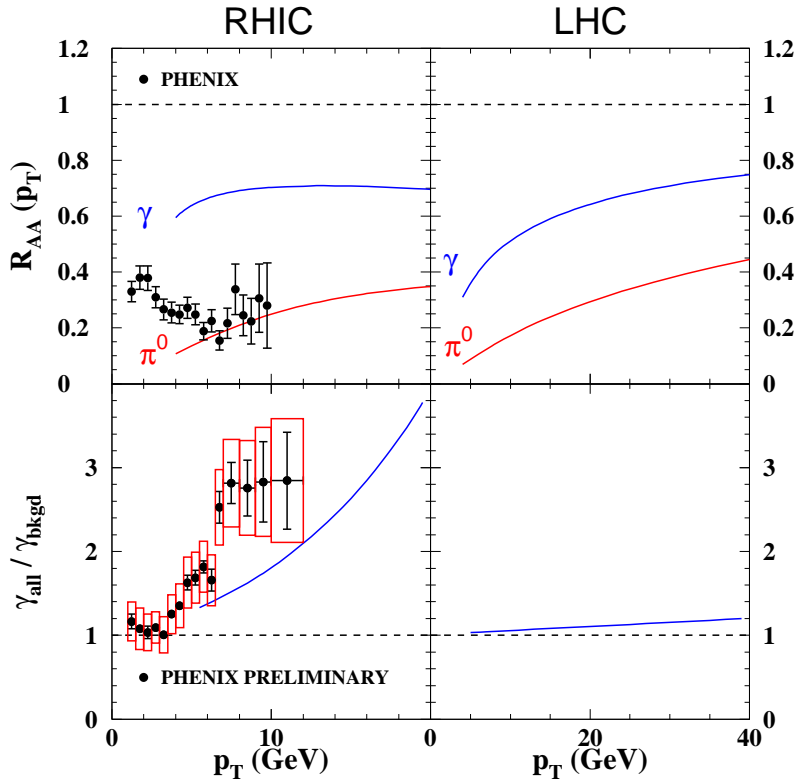,width=11.0cm}}
\vspace{-0.2cm}
\caption{{\it Top}: Quenching factor for $\gamma$ and $\pi^0$ at RHIC ({\it left}) and LHC ({\it right}). {\it Bottom}: Photon signal over background ratio at RHIC ({\it left}) and LHC ({\it right}).}
\end{figure}
\end{center}

In Figure~\ref{fig:rhiclhc} ({\it upper panel}) is computed the quenching factor, namely the ratio of $p_{_T}$ spectra in Au-Au over those in $p$-$p$ collisions for both photons and pions. At RHIC energy ({\it upper left}), the $\pi^0$ attenuation measured by the PHENIX collaboration~\cite{Adcox:2001jp} is well described by the perturbative calculation, although these clearly underestimate the data at very low $p_{_T} \simeq 4$~GeV. At such low scales, however, the picture of a hard probe penetrating the soft medium is no longer relevant (see Introduction) and many other processes may compete, such as thermal production. Photon production is much less suppressed than $\pi^0$ from the rather large direct component (Fig.~1, {\it left}) at RHIC energy. The photon quenching seems to flatten out and to remain large at high $p_{_T}$. Nevertheless, let us mention that this actually comes from the interplay between energy loss effects and significant isospin corrections which tend to suppress the photon yield in an (almost) isoscalar Au nucleus as compared to a proton, $u_{_{Au}}(x) \ll u_{_p}(x)$. Similar features are observed when going from RHIC to LHC energy ({\it upper right}), although the isospin effect is now negligible from the smaller $x$ partons (gluon dominated) probed in the nuclei.

Low transverse momentum photons are difficult to identify at RHIC because of the large background coming from the $\pi^0$ radiative decays. Nevertheless, the total photon yield normalized to the background has been extracted recently by the PHENIX experiment~\cite{Frantz:2004gg} and is shown in Figure~\ref{fig:rhiclhc} ({\it lower left}). Although the trend of the calculation looks similar to that of the data, it is worth noting that this ratio is systematically underestimated. The reason is not yet clear. This photon ``excess'' may be due to the medium-induced photon radiation from the leading parton which could be important~\footnote{Note that such a contribution is not taken into account in our calculation.} when the transverse momentum is not too large, as recently suggested by Zakharov~\cite{Zakharov:2004bi}.

At LHC, the $p_{_T}$ dependence of the ratio is much less pronounced than at RHIC energy. Indeed, the huge gluon luminosity at this energy favors $\pi^0$ production as compared to $\gamma$ and thus makes the ratio much closer to one. Moreover, the photon yield will be more affected by the energy loss process than at RHIC from the larger fragmentation component, which decreases the ratio accordingly.

\section{Diphoton correlations at LHC}

Single inclusive photon and pion $p_{_T}$ spectra discussed in the previous section proved sensitive to the parton energy loss in the dense QCD medium. However, although  quantitative effects are observed at RHIC and LHC such observables do not allow to study what the kinematic dependence of this process is. In particular, the fragmentation functions modified by the medium are not directly accessible through these measurements as the initial parton energy $k_{_T}$ --~hence the momentum fraction $z$ carried away by the photon (pion)~-- cannot be determined. Hopefully, $\gamma$-tagged measurements could help with in this respect~\cite{Wang:1996yh}. 

\begin{center}
\begin{figure}[htbp]\label{fig:lhc}
\centerline{\psfig{figure=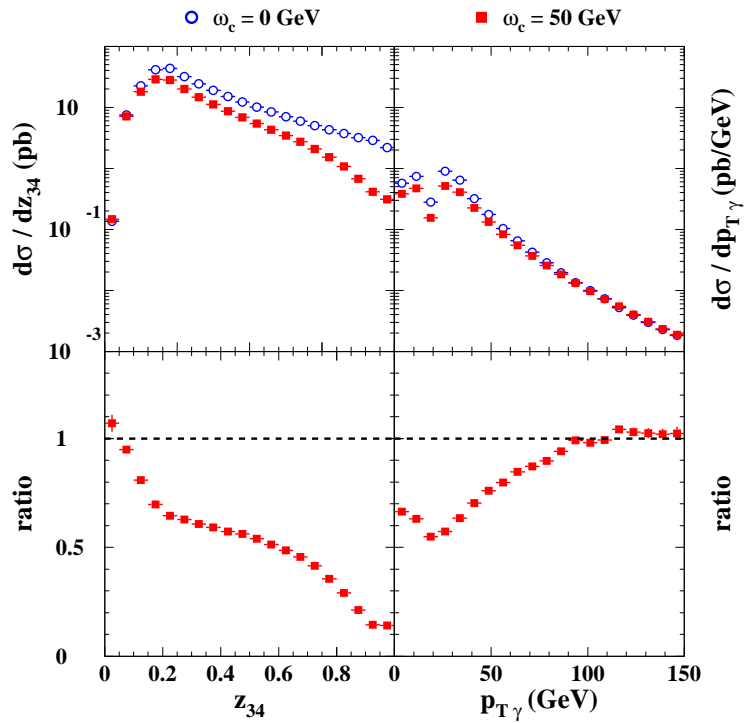,width=10.cm}}
\vspace{-0.2cm}
\caption{{\it Top}: Diphoton production cross section at LHC as a function of $z_{3 4}$ ({\it left}) and $p_{_T}$ ({\it right}). {\it Bottom}: Ratios of the $\omega_c = 50$~GeV over $\omega_c = 0$~GeV spectra.}
\end{figure}
\end{center}

Therefore, we determine in this section the diphoton production cross section at LHC to leading order~\footnote{Perturbative calculations were performed using the work Ref.~\cite{Binoth:1999qq}.} using the the medium-modified fragmentation functions~(\ref{eq:modelFF}). Asymmetric cuts on the photon transverse momenta were applied, $p_{_{T 1}} \ge 25$~GeV, $p_{_{T 2}} \ge 5$~GeV to favor the 1-fragmentation dynamical component, with one photon produced directly and one coming from the parton fragmentation~\footnote{On top of the 2-direct and 1-fragmentation component (Fig.~\ref{fig:components}), the two photons may result from the collinear fragmentation of the two back-to-back partons at leading order.}. Figure~\ref{fig:lhc} ({\it upper left}) displays the diphoton production cross section as a function of the momentum imbalance
\[
z_{3 4} = - \frac{{\vec p_{_{T 1}}}\,.\,{\vec p_{_{T 2}}}}{p^2_{_{T 1}}}.
\]
for $\omega_c = 0$ and 50~GeV. Assuming the 1-fragmentation dominates, we have at leading order $z_{3 4} \simeq z$. The distribution $d\sigma/dz_{3 4}$ is thus reminiscent of the photon fragmentation function, $D_{\gamma / k}(z_{3 4})$. This observable offers therefore a ``direct'' access to the medium-modified fragmentation functions. In particular, we observe ({\it lower left}) a strong depletion of the quenching factor as $z_{3 4}$ gets close to 1, as expected from the restricted phase space Eq.~(\ref{eq:modelFF}). Such a behavior is also observed in hadron production in semi-inclusive DIS data on nuclei~\cite{Arleo:2003jz}.

The photon $p_{_T}$ spectra are also determined ({\it upper right}). Within that kinematics, the cross section shows a discontinuity since only one of the two photons is measured below the upper cut, $p_{_{T 2}} \le 25$~GeV. Looking at the ratio of $p_{_T}$ spectra ({\it lower right}), the quenching is maximal around the upper cut. Indeed, as $p_{_T}$ approaches the upper cut from ``below'', events with larger $z$ are selected, $p_{_{T 1}} \simeq p_{_{T 2}}$, where energy loss effects are most pronounced. Above that cut, the proportion of directly produced photons (unaffected by the medium) is getting larger and the quenching factor is slowly reaching unity as expected at aymptotic energies, $p_{_T} \gg \omega_c$.
\vspace{-0.2cm}
\vspace{-0.58cm}
\noindent \emph{Acknowledgements.} Part of this work has been done in collaboration with P.~Aurenche and Z.~Belghobsi. I also thank H.~Delagrange for useful comments.

\end{document}